\documentclass[final,3p,twocolumn,times,11pt]{elsarticle}
\usepackage{amssymb}
\usepackage{amsmath}
\usepackage{color}
\usepackage{xspace}
\usepackage{multirow}
\usepackage[utf8x]{inputenc}
\graphicspath{{./}}

\journal{Physics Letters B}

\begin{document}

\begin{frontmatter}

\title{Heavy meson tomography of cold nuclear matter at the electron-ion collider}

\author{Hai Tao Li}
\ead{haitaoli@lanl.gov}
\author{Ze Long Liu}
\ead{zelongliu@lanl.gov}
\author{Ivan Vitev}
\ead{ivitev@lanl.gov}
\address{Theoretical Division, Los Alamos National Laboratory, Los Alamos, NM, 87545, USA}

\begin{abstract}
An important part of the physics program of the future electron-ion collider is to understand the nature of hadronization and the transport of energy and matter in large nuclei. Open heavy flavor production in deep inelastic scattering  provides a new tool to address these critical questions.  We present the first calculation of  $D$-mesons and $B$-meson cross sections in electron-nucleus collisions at the EIC by including both next-to-leading order QCD corrections and cold nuclear matter effects. Our formalism employs generalized DGLAP evolution  to include the contribution of in-medium parton showers, and is based on  methods developed in soft-collinear effective theory with Glauber gluons that describe inclusive hadron production in reactions with nucleons and  nuclei. The comprehensive study summarized here allows us to identify the optimal observables, center-of-mass energies, and kinematic regions most sensitive to the physics of energy loss and hadronization at the EIC.  
\end{abstract}   

\end{frontmatter}

\section{Introduction} 

 A  high-luminosity  electron-ion collider (EIC), which recently received mission need approval from the US Department of Energy,  can address fundamental questions about nucleons  and nuclei. These include the origin of mass, the internal landscape of hadrons, the phenomenon of gluon saturation, and the physics of hadronization~\cite{Aidala:2020mzt}.   Production and propagation of heavy flavor in deep inelastic scattering (DIS) is a unique and critical part of this planned decade-long research program.  Studies in this direction have so far focused on charm production 
 that can constrain the gluon and strangeness content of the nucleon/nucleus~\cite{Chudakov:2016otl,Chudakov:2016ytj,Arratia:2020azl}, especially at moderate and high values of Bjorken-$x$.

 In this letter we investigate semi-inclusive open heavy meson cross sections in electron-proton (e+p) and electron-nucleus (e+A) collisions to address a different set of questions -  how energy and matter are transported through a strongly interacting quantum mechanical environment.  The possibility to study the physics of hadronization and energy loss of partons in cold nuclear matter has been investigated by the HERMES Collaboration at HERA using fixed nuclear targets and an electron beam of energy $E_{\rm beam} = 27.6$~GeV~\cite{Airapetian:2000ks, Airapetian:2003mi,Airapetian:2007vu}.  In these experiments suppression of the multiplicities of light hadrons  in e+A  versus e+p  collisions has been  clearly established.   Hadronization in nuclei was also studied  experimentally  earlier in Refs.~\cite{Osborne:1978ai,Ashman:1991cx,Adams:1993mu}.  Theoretical interpretations of the data predominantly fall within two classes of models. The parton energy loss approach assumes that  fragmentation occurs outside of the nucleus and evaluates the attenuation of the  
quarks and gluons that produce the final-state hadrons, or, equivalently, the effective modification of fragmentation as a function of the transport properties of large 
nuclei~\cite{Wang:2002ri,Arleo:2003jz,Chang:2014fba}. The hadron absorption model argues that hadronization can take place on length scales smaller than the nuclear size and the final-state particle can be absorbed in the medium~\cite{Accardi:2002tv,Kopeliovich:2003py}. We  note that phenomenology that uses elements of both elastic parton scattering and hadron absorption has been developed~\cite{Brooks:2020fmf}. Transport models have also been employed to investigate the HERMES data~\cite{Falter:2004uc}.  For a review of energy loss and hadronization in cold nuclear matter, see Ref.~\cite{Accardi:2009qv}.

The HERMES Collaboration e+A results have advanced our understanding of particle production in the nuclear environment, but a number of open questions still remain. The transport coefficients extracted from data using different energy loss approaches differ by up to an order of magnitude~\cite{Arleo:2003jz,Chang:2014fba}. More importantly, fundamentally different assumptions about the time scales involved in the process of hadronization and the nature of nuclear attenuation - inelastic parton scattering versus hadron absorption - give equally good description of  the light meson multiplicities' quenching~\cite{Chang:2014fba,Kopeliovich:2003py}.  With this in mind, we turn to open heavy meson production as a new probe of cold nuclear matter effects at the EIC, where the semi-inclusive cross sections can be readily measured. Since the shape of charm and beauty quark fragmentation functions (FFs) into   $D$-mesons and $B$-mesons is very different  from the shape of light parton fragmentation into pions and kaons, carefully chosen observables may be much more sensitive to the nature of nuclear attenuation~\cite{Li:2020sru}.  A number of center-of-mass (CM) energies are expected to be available at the EIC, with multiple distinct kinematic domains for the final-state  particles   for each electron-proton/nucleon energy combination. Thus, we further aim to identify the CM energies and  rapidity intervals  that are most sensitive to the nuclear modification of hadron production from final-state interactions,  which may facilitate operation planning and optimize detector coverage for the EIC.

In describing heavy meson production in DIS on nuclei we go beyond the traditional energy loss approach.
The evolution of FFs  is determined by the Dokshitzer-Gribov-Lipatov-Altarelli-Parisi (DGLAP) evolution equations. Recently, soft-collinear effective theory  (SCET)~\cite{Bauer:2000yr,Bauer:2001yt,Bauer:2002nz,Beneke:2002ph} has been employed to describe interactions between jets and a QCD medium via Glauber gluon exchange~\cite{Idilbi:2008vm,Ovanesyan:2011xy}.  This allowed for the derivation of the full set of $1\to 2$ medium-induced splitting kernels for massless and massive partons~\cite{Ovanesyan:2011kn,Kang:2016ofv}.  With EIC applications in mind, these results were verified using a lightcone wavefunction approach and a formalism to calculate higher order corrections in the opacity of nuclear matter was developed~\cite{Sievert:2018imd,Sievert:2019cwq}.    The medium-induced splitting kernels can be used to understand the evolution of FFs in cold nuclear matter - a technique which was successfully developed in heavy ion collisions~\cite{Kang:2014xsa,Chien:2015vja}. In this letter, we employ  the QCD evolution-based method to encode the cold nuclear matter effects on hadron production at the EIC and to present the results of our analysis.

The rest of our letter is organized as follows. In section 2, we briefly introduce the theoretical framework for the next-to-leading order (NLO) QCD corrections to  hadron production in DIS and in-medium QCD evolution based on SCET$_{\rm G}$. In section 3, we compare theoretical predictions with HERMES measurement and demonstrate the validity of our approach for hadron production in reactions with heavy nuclei. Section 4 is dedicated to the detailed study of pion, $D$-meson and $B$-meson production at the EIC. We conclude in section 5.

\section{Theoretical Framework}

\begin{table*}[!t]
	\begin{center}
		\begin{tabular}{c|c|c|c|c|c|c|c} 
			\hline			
			\hline 
			\multicolumn{2}{c|}{Energy}
			& \multicolumn{2}{c|}{5 GeV$\times$40 GeV}
			& \multicolumn{2}{c|}{10 GeV$\times$100 GeV}
			& \multicolumn{2}{c}{18 GeV$\times$275 GeV} \\
			\hline
			\multicolumn{2}{c|}{$p_T^h$ [GeV]} &~ [2\,,3]~&~ [5\,,6] ~
			&~ [2\,,3]~&~ [5\,,6] ~
			&~ [2\,,3]~&~ [5\,,6] ~\\
			\hline
			\multirow{2}{*}{$\pi^+$}  
			& LO
			&  $5.3 \times 10^{6} $  &  $2.4  \times 10^{4} $
			&   $ 1.4\times 10^{7} $ &  $3.0 \times 10^{5} $ 
			&   $2.9 \times 10^{7} $ &  $9.6 \times 10^{5} $\\
			& NLO
			&  $1.1 \times 10^{7} $ &   $6.9 \times 10^{4}  $
			&   $2.8 \times 10^{7} $ &  $ 6.1\times 10^{5} $ 
			&  $ 5.6\times 10^{7} $  &  $1.9\times 10^{6} $\\
			\hline
			\multirow{2}{*}{$D^0$}  
			& LO
			& $ 1.4\times 10^{6} $  &   $3.2 \times 10^{3}  $
			& $ 8.6\times 10^{6} $   &  $9.0 \times 10^{4}  $
			& $ 3.1\times 10^{7} $   &  $6.6 \times 10^{5} $\\
			& NLO
			& $ 3.7\times 10^{6} $ &   $8.5 \times 10^{3}  $
			& $ 2.1\times 10^{7} $   & $ 2.1\times 10^{5} $
			&  $ 7.2\times 10^{7} $  & $1.5 \times 10^{6} $ \\
			\hline			
			\multirow{2}{*}{$B^0$}  
			& LO
			& $ 3.7\times 10^{5}$   &   $1.2 \times 10^{3}  $
			&  $2.4 \times 10^{6} $   & $2.8 \times 10^{4}  $
			&  $9.0 \times 10^{6} $   & $ 2.0\times 10^{5} $ \\
			& NLO
			& $ 1.1\times 10^{6}$  &   $3.3 \times 10^{3}  $
			&  $6.2 \times 10^{6} $   & $7.2 \times 10^{4}  $
			&  $ 2.1\times 10^{7} $   & $ 4.7\times 10^{5} $ \\
			\hline			
			\hline			
		\end{tabular}
		\caption{\label{tab:eventnum}  Example of light, charm, and bottom hadron multiplicities at the EIC in selected $p_T$ bins ($2\ {\rm GeV}<  p_T^h< 3\ {\rm GeV}$ and $5\ {\rm GeV}< p_T^h< 6\ {\rm GeV}$) to lowest and next-to-leading order.  We have integrated over the hadron rapidity in the interval $-2<\eta <4$ and  used a typical      one year integrated luminosity  of $10\,{\rm fb}^{-1}$ in e+p collisions. 
		}
	\end{center}
\end{table*}

\subsection{Hadron Production in DIS }
In collinear leading-twist perturbative QCD the inclusive cross section for the production of hadron  $h$  is factorized  as follows:
\begin{equation}\label{eq:NLOform}
\begin{aligned}
E_{h} &\frac{d^{3} \sigma^{\ell N \rightarrow h X}}{d^{3} P_{h}} 
=\frac{1}{S} \sum_{i, f} \int_{0}^{1} \frac{d x}{x} \int_{0}^{1} \frac{d z}{z^{2}} f^{i / N}(x, \mu)\\
& \times D^{h / f}(z, \mu) \Big[\hat{\sigma}^{i \rightarrow f}
+f_{\rm ren}^{\gamma /\ell}\left(\frac{-t}{s+u},\mu\right)\hat{\sigma}^{\gamma i \to f}\Big] \, .
\end{aligned}
\end{equation}
Here,  $ f^{i / N}$ is the parton distribution function (PDF) of parton $i$ in nucleon $N$ and $D^{h / f}$ is the conventional FF from parton $f$ to hadron $h$.  $\hat{\sigma}^{i\to f}$ is the partonic cross section for lepton-parton scattering with initial-state parton $i$  and final-state parton $f$.  $s$, $t$, $u$ are the partonic Mandelstam variables defined as $s=(k+l)^2$, $t=(k-p)^2$ and $u=(l-p)^2$, where $l^\mu$, $k^\mu$ and $p^\mu$ are the momenta of incoming lepton, incoming parton and fragmenting parton, repectively.
 In hadron and jet production at the EIC it is not necessary to place kinematic constraints on the scattered lepton. Thus,  events with  lepton scattering at a small angle can be selected. Then, the hard process can be described by an incoming quasi-real photon scattering: $\gamma q\to q(g)$, $\gamma q\to g(q)$, $\gamma g\to q({\bar q})$, which contribute to the cross section starting at order $\alpha_{\rm EM}^2 \alpha_s$. 
In this case, the incoming lepton is regarded as a source of quasi-real photons. The well known Weizs\"acker-Williams (WW) distribution provides an accurate description for photons in leptons by a perturbative distribution function $f_{\rm ren}^{\gamma /\ell}\left(y,\mu\right)$~\cite{vonWeizsacker:1934nji,Williams:1934ad,Bawa:1989bf,Frixione:1993yw}. The analytical expressions for  $\hat{\sigma}^{i\to f}$, $\hat{\sigma}^{\gamma i\to f}$ and $f_{\rm ren}^{\gamma /\ell}\left(y,\mu\right)$  have been known up to ${\cal O}(\alpha_{\rm EM}^2 \alpha_s)$ for a while, and can be found in~\cite{Hinderer:2015hra}.

In the numerical calculations that follow we  use CT10nlo PDF sets~\cite{Lai:2010vv} 
and the associated strong coupling provided by  {\sc  Lhapdf6}~\cite{Buckley:2014ana}. Fragmentation functions into light hadrons, for example 
$\pi$, are taken from  Ref.~\cite{Hirai:2007cx}.   The boundary condition for heavy  quark fragmentation into the various   $D$-meson and $B$-meson states at a scale $\mu = 2 m_Q$ can be calculated perturbatively using heavy quark effective theory (HQET), as shown in Refs.~\cite{Braaten:1994bz,Cheung:1995ye}.  The FFs obey the DGLAP evolution equations, which can be written as   
\begin{multline}~\label{eq:dglap}
    \frac{d}{d \ln \mu^2} D^{h/i}\left(x, \mu\right)=
    \\
    \sum_{j} \int_{x}^{1} \frac{d z}{z} P_{j i}\left(z, \alpha_{\mathrm{s}}\left(\mu\right)\right) 
  D^{h/j}\left(\frac{x}{z}, \mu\right) \, , 
\end{multline}
where $P_{j i}$ is the Altarelli-Parisi (AP) splitting functions describing $i\to j + X $ splitting and $z$ is the  longitudinal momentum fraction of $j$ relative to $i$.   
We take the perturbative hard part at NLO and use PDFs consistent with NLO global analysis. The evolution of fragmentation functions is at one loop, because the medium corrections to the splitting functions in Eq.~(5) are only available at LO~\footnote{An exploratory study of  the real contribution to higher order parton splitting in matter was carried out in~\cite{Fickinger:2013xwa}, but the result is complex and its numerical evaluation  challenging with current and near-future computing resources to be practically applicable to phenomenology.}. We employ the LO splitting functions in vacuum as well for consistency and
make use of {\sc Hoppet}~\cite{Salam:2008qg} to solve  Eq.~(\ref{eq:dglap}) numerically.

To understand the feasibility of heavy flavor measurements at nominal EIC luminosity  and to assess the magnitude of higher order corrections we first turn to the calculation  of hadron cross sections in e+p collisions. The vacuum splitting functions are used to perform the RG evolution of the FFs. Both the renormalization scale and factorization scale are chosen as the energy of the initial parton fragmenting to a  hadron  in the rest frame of the proton.  This is motivated by the need for consistency with e+A calculations where the energy of the parent quark or gluon in nuclear matter plays a key role in determining the strength of the medium-induced parton shower. 	
Selected results for the  expected multiplicities  of light, charm, and beauty mesons,  exemplified by $\pi^+$, $D^0$ and $B^0$, are shown in  Table~\ref{tab:eventnum} for integrated luminosity of  10~fb$^{-1}$. We consider three combinations of electron and proton beam energies: 
5~GeV (e) $\times$ 40 GeV (p), 10~GeV (e) $\times$ 100 GeV (p), and 10~GeV (e) $\times$ 100 GeV (p)  and integrate over the rapidity interval  $-2 < \eta < 4$.
 The NLO QCD corrections are obtained from Eq.~(\ref{eq:NLOform}), including the contribution from quasi-real photon scattering.  They lead to a $K$-factors  in the range of 1.5 to 2.5. For $\pi$ meson production, the quasi-real photon scattering contributes about 40\% to 50\% to NLO corrections, while for $D$-mesons and $B$-mesons the quasi-real photon contribution is even more dominant. 
The NLO corrections are sizable and when it comes to absolute cross sections they should be considered for reliable theoretical predictions.

\subsection{Cold Nuclear Matter Effects }

\begin{figure}[!t]
    \centering
    \includegraphics[scale=0.47]{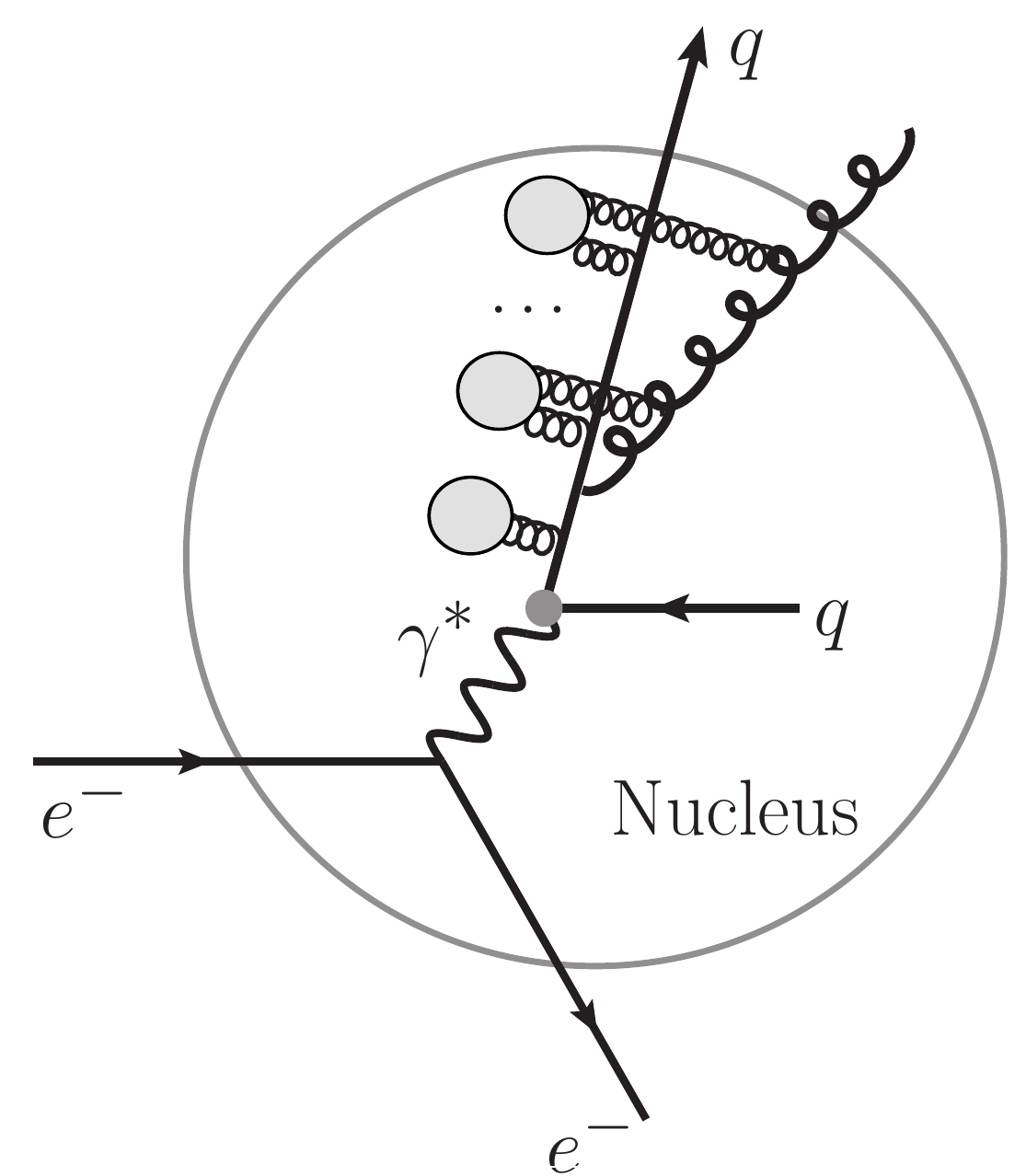}
    \caption{Illustration of in-medium parton shower formation in electron-nucleus collisions from the interactions of the struck quark. It will affect the evolution 
    of fragmentation functions and, ultimately, the cross sections for light and heavy hadron production. }
    \label{fig:EIC}
\end{figure}

When partons propagate in strongly-interacting matter they scatter and radiate. The medium-induced parton shower  will modify the evolution of the FFs,  and has been investigated in the framework of SCET$_{\rm (M), G}$~\cite{Ovanesyan:2011xy,Kang:2016ofv}.  
These  modifications were first introduced as  corrections to the DIS hadronization process~\cite{Wang:2001ifa}  and, more recently, implemented  in medium-modified DGLAP evolution. This theoretical framework has been used extensively in Refs.~\cite{Chang:2014fba,Kang:2014xsa,Chien:2015vja,Kang:2016ofv,Wang:2009qb,Li:2017wwc, Li:2019dre} to carry out resummation  in cold and hot QCD medium numerically,  and to describe hadron production and observables sensitive to the fragmentation process.  We will solve the medium-corrected DGLAP evolution equations to take account of the  radiation induced by a large nucleus, as shown in Fig.~\ref{fig:EIC}. 

The full  fragmentation function evolution in the presence of nuclear matter is  given by:
\begin{multline} \label{eq:fullevol}
    \frac{d}{d \ln \mu^{2}} \tilde{D}^{h/i}\left(x, \mu\right)= 
    \sum_{j} \int_{x}^{1} \frac{d z}{z}  \tilde{D}^{h/j}\left(\frac{x}{z}, \mu\right)  
    \\ \times
   \left( P_{j i}\left(z, \alpha_{s}\left(\mu\right)\right) +  P_{j i}^{\rm med}\left(z, \mu\right)  \right)  \, .
\end{multline}
In Eq.~(\ref{eq:fullevol})  $ P_{j i}^{\rm med}$ are the medium corrections to the splitting functions. It has been demonstrated   
that the full splitting kernel is a direct sum of its vacuum and medium-induced components and  the corrections are 
gauge-invariant. 
We will make use of the form of  in-medium branching  processes derived in~\cite{Ovanesyan:2011xy,Ovanesyan:2011kn,Sievert:2018imd,Sievert:2019cwq}.  Equivalent to the vacuum splitting functions, the real contribution  can be written as
\begin{align}
    P_{ j i }^{\mathrm{med},\rm{real}} \left(z, \mathbf{k}_{\perp}\right) = 2\pi\, \mathbf{k}_{\perp}^2  \frac{dN_{j i }^{\mathrm{med}} }{d^2\mathbf{k}_{\perp}  dz} \,.
\end{align} 
The full splitting functions can be expressed as proportional to the vacuum ones with a medium induced correction that depends both on the longitudinal momentum fraction $z$ and the intrinsic transverse momentum of the branching $\mathbf{k}_{\perp}$.   This is because in-medium parton showers are broader and softer than the ones in the vacuum.

The full set of medium corrections to the splitting functions can be written as
\begin{align}\label{eq:sp}
    P_{qq}^{\rm{med}}\left(z, \mathbf{k}_{\perp}\right)&=\left[P_{q \rightarrow q g}^{\mathrm{med},\rm{real}} \left(z, \mathbf{k}_{\perp}\right)\right]_{+} \; ,
    \nonumber \\ 
    P_{gq}^{\rm{med}}\left(z, \mathbf{k}_{\perp}\right)  &=  
       P_{q \rightarrow g q }^{\mathrm{med},\rm{real}} \left(z, \mathbf{k}_{\perp}\right) \; ,
    \nonumber \\
     P_{qg}^{\rm{med}}\left(z, \mathbf{k}_{\perp}\right)  &=  
        P_{g\rightarrow  q\bar{q}}^{\mathrm{med},\rm{real}} \left(z, \mathbf{k}_{\perp}\right) \; ,
    \nonumber \\ 
    P_{gg}^{\rm{med}}\left(z, \mathbf{k}_{\perp}\right)  &= 
     \left[ \left(\frac{2z-1}{1-z}+z(1-z) \right) h_{gg} \left(z, \mathbf{k}_{\perp}\right) \right]_+ 
     \nonumber \\ &
     + \frac{ h_{gg} \left(z, \mathbf{k}_{\perp}\right) }{z} + B(\mathbf{k}_{\perp}) \delta(1-z) \; ,
\end{align}
where 
\begin{align}
    h_{gg} \left(z, \mathbf{k}_{\perp}\right)  = & \frac{ P_{g \rightarrow gg }^{\mathrm{med},\rm{real}}
   \left(z, \mathbf{k}_{\perp}\right)}{ \frac{z}{1-z} + \frac{1-z}{z}+z(1-z)}  \; ,
\end{align} 
and $B(\mathbf{k}_{\perp})$ can be obtained through momentum sum rules. The definition of the splitting function in QCD medium can be also found in Refs.~\cite{Kang:2014xsa,Chien:2015vja}. 
Going back to Eq.~(\ref{eq:fullevol}), $k_\perp$ which characterizes the intrinsic momentum of the collinear branching is the scale we chose for the medium-induced splitting functions. 
The evolution for heavy flavor can  similarly be written down 
and the splitting kernels associated with massive quarks can be found in Refs.~\cite{Kang:2016ofv,Sievert:2019cwq}.  
The medium-induced splitting functions for massive quarks are defined in a similar way as the ones in Eq.~(\ref{eq:sp}). They reduce 
to the massless case for large momentum scales, while the mass effects can play an important role for small momentum scales. 

\begin{figure}
	\centering
	\includegraphics[scale=0.55]{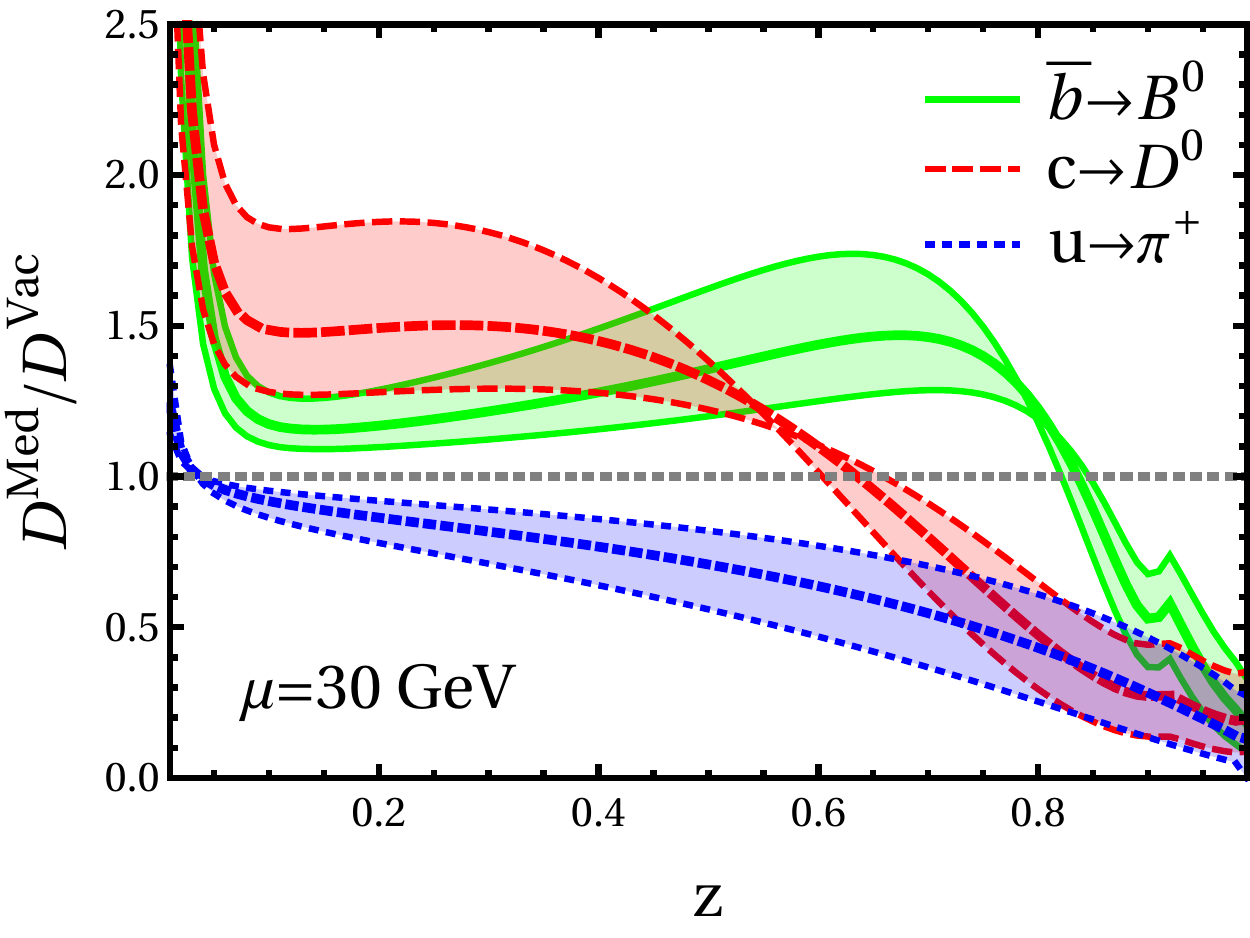}
	\caption{The ratio of fragmentation functions for the case of a Au nucleus to the ones in vacuum at a scale $\mu$ = 30 GeV. 
	Blue band (dotted lines), red band (dashed lines), and green band (solid lines) correspond to light parton to pion, $c$-quark
	to $D$-meson, and  $b$-quark to  $B$-meson fragmentation, respectively.
	 }
	\label{fig:FFsInMedium}
\end{figure}

\begin{figure*}[!t]
	\centering
	\includegraphics[width=0.42\textwidth]{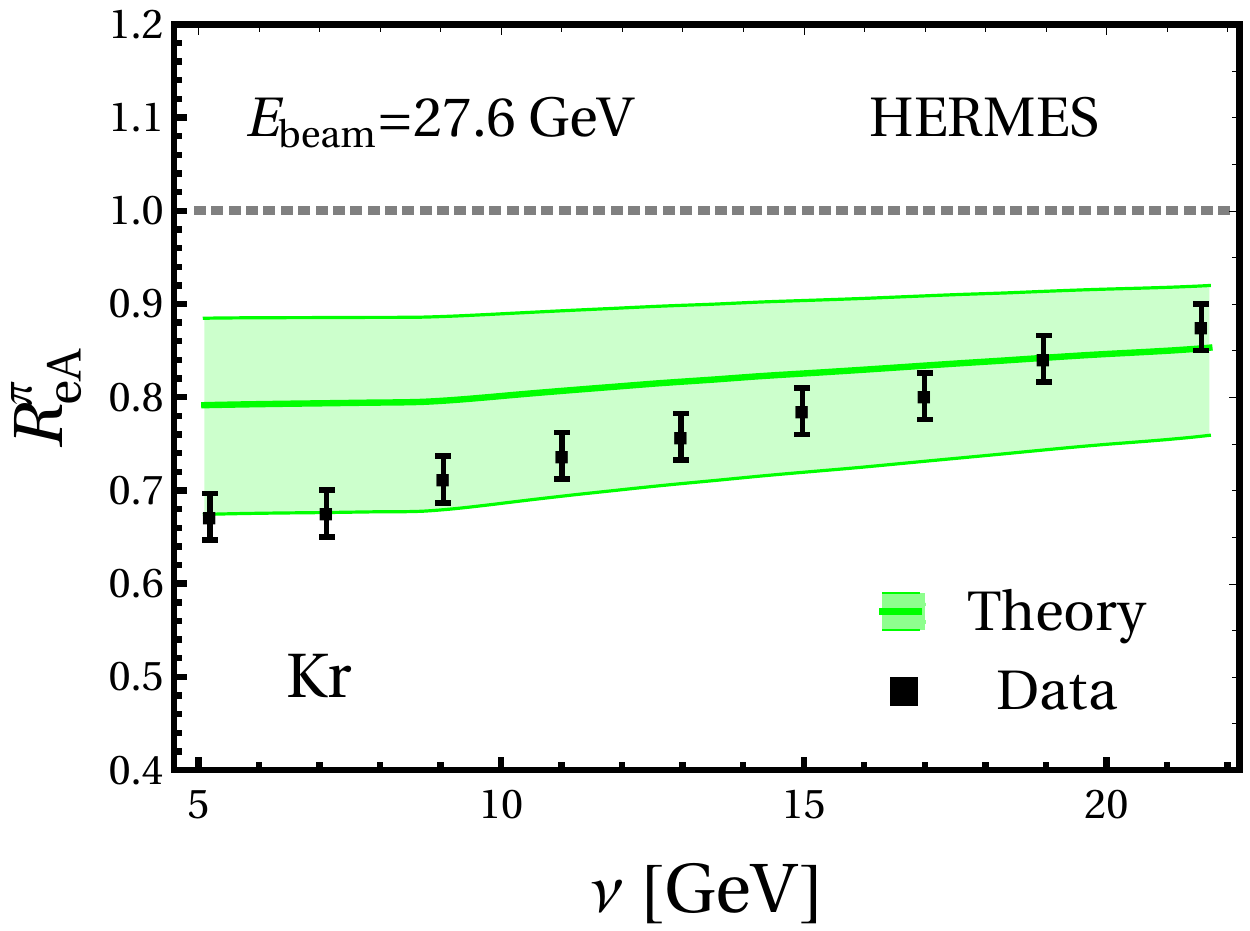}\quad
	\includegraphics[width=0.42\textwidth]{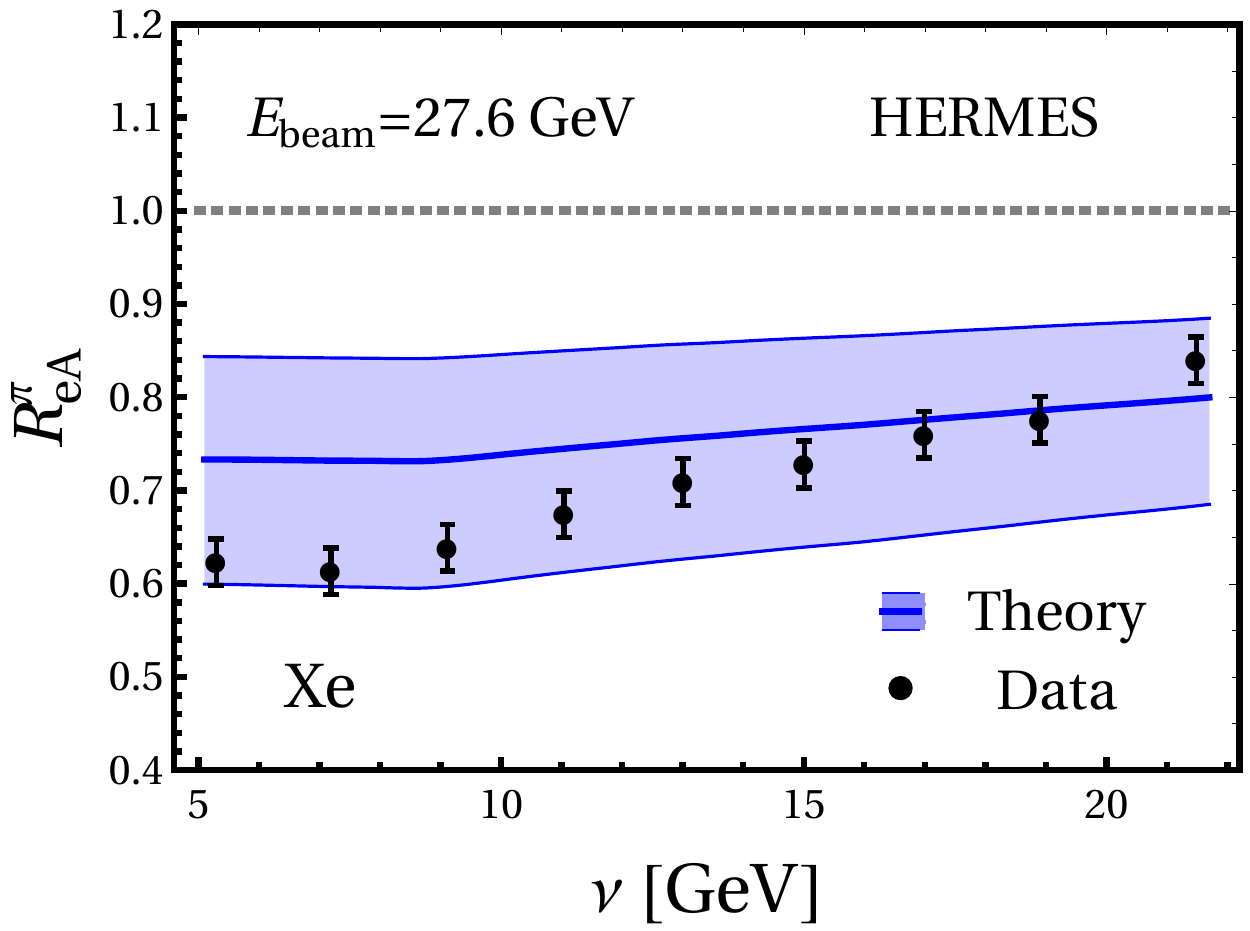}       \\[2ex]
	\includegraphics[width=0.42\textwidth]{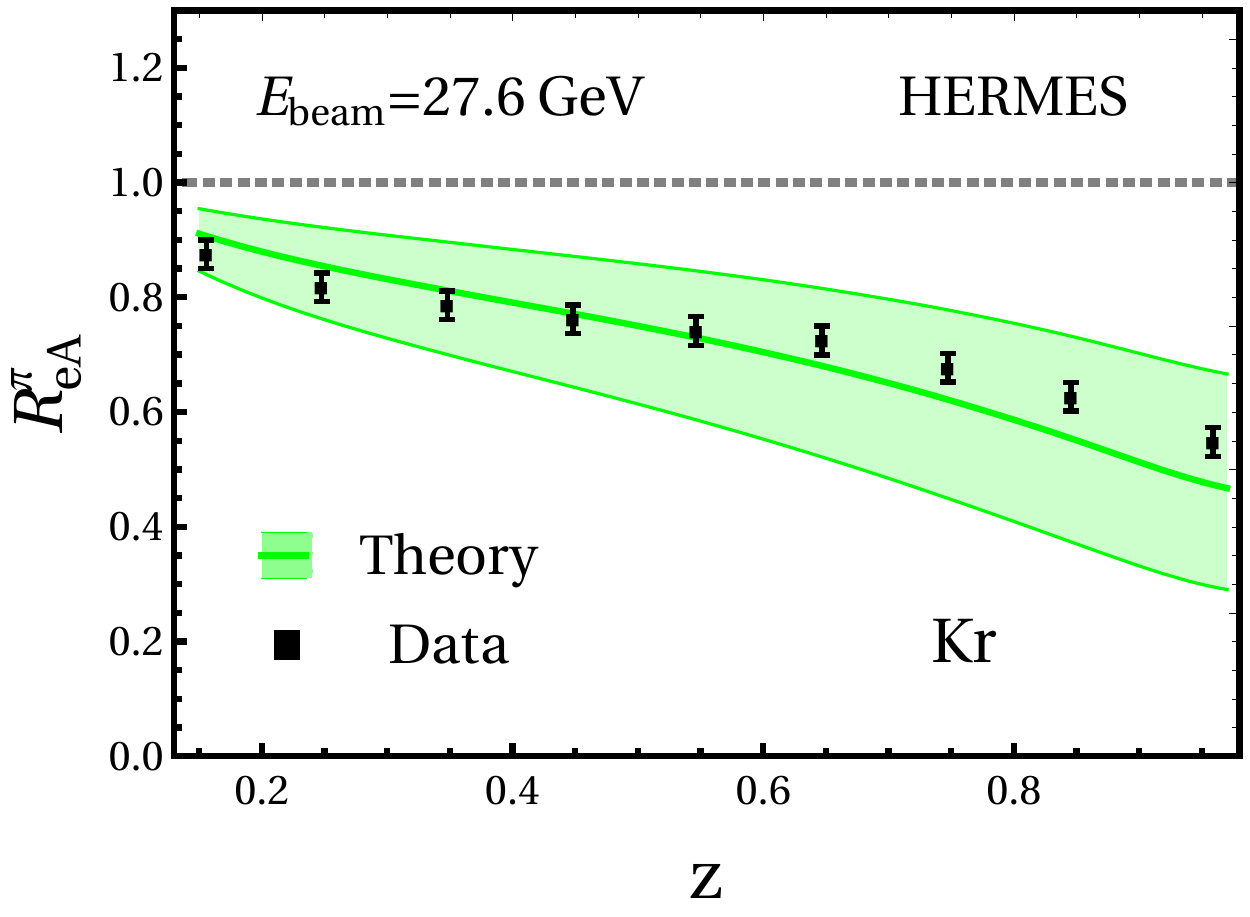}\quad
	\includegraphics[width=0.42\textwidth]{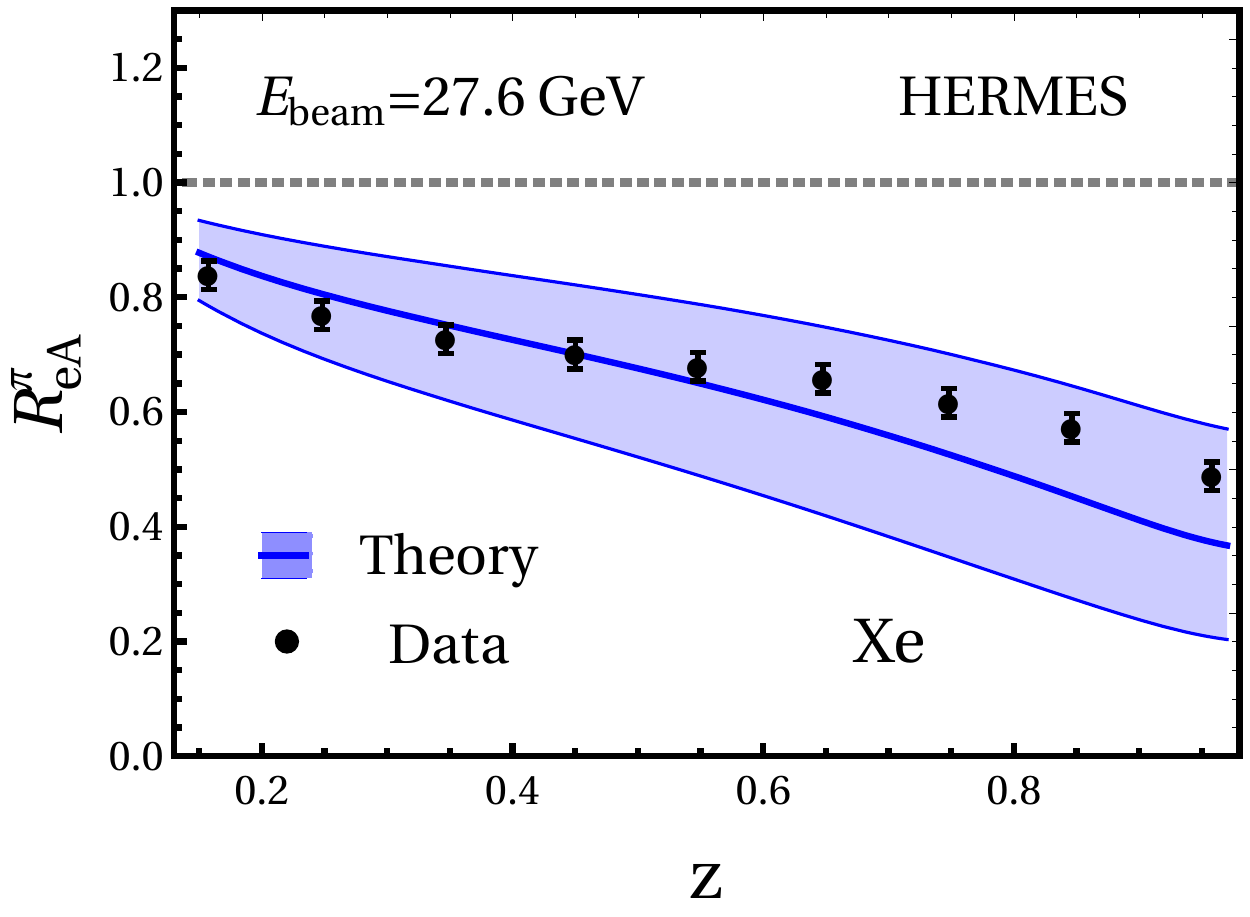}   	
	\caption{ Top panels: comparison of $R_{eA}$ for $\pi^+$ as a function of the energy  $\nu$ with HERMES measurements~\cite{Airapetian:2007vu}. The bands correspond to a variation in the transport properties  of cold nuclear matter. Bottom panels: similar comparison, but as a function of the fragmentation fraction $z$.
	Left panels are for the Kr target and right panels are for the Xe target, respectively.}
	\label{fig:hermes}
\end{figure*} 

An example of how in-medium evolution can alter the fragmentation pattern of partons into hadrons is given in Fig.~\ref{fig:FFsInMedium}.
It presents the ratio of the  FFs for the case of a gold (Au)  nucleus evolved from the boundary condition to a scale $\mu=30$~GeV 
to the ones in the vacuum, denoted $D^{\rm Med}/D^{\rm Vac}$.  The dotted blue lines, dashed red lines and solid green lines represent the fragmentation of $u\to \pi^+$, $c\to D^0$ and ${\bar b}\to B^0$, respectively.  We have averaged the parent parton production  point over the nuclear geometry in evaluating  the splitting kernels that enter the evolution equations.   The nominal transport coefficient of cold nuclear matter, which are determined by HERMES data as shown in Fig.~\ref{fig:hermes},  we take to be $ \langle q^2_\perp \rangle / \lambda_g = 0.12 $~GeV$^2$/fm for gluons and $ \langle q^2_\perp \rangle / \lambda_q = 0.05 $~GeV$^2$/fm for quarks. Here,  $\langle q^2_\perp \rangle $  is the mean momentum transfer squared in two dimensions per scattering and $\lambda_g $ ($\lambda_q $) are the gluon (quark) scattering lengths, respectively.  The bands correspond to varying the transport parameter up and down by a factor of two.   
The effect of the medium-induced shower is to further soften fragmentation relative to the vacuum. We can see that the FFs for $\pi^+$ are always suppressed,  except for very small values of $z$. The fragmentation pattern of heavy flavor is modified in a distinctly different way, the suppression only happens in the large-$z$ region. For  $c\to D^0$ and ${\bar b}\to B^0$, the in-medium corrections enhance very significantly FFs with $z<0.6$ and $z<0.85$, respectively.  
In addition, the modification due to cold nuclear matter effects is larger at lower energy scales, which opens the door toward fruitful phenomenology at the future EIC. 
An essential task that we face is to identify the optimal phase space regions that are most sensitive  to 
the effect of in-medium parton showers  and where semi-inclusive DIS measurements can provide constraints on the transport properties of large nuclei.

Lastly, we remark that the in-medium corrections to FFs for $c\to D^0$ and ${\bar b}\to B^0$ rise  for very small values of $z \rightarrow 0$. The physical reason for this behavior is that in-medium evolution produces even more soft partons than vacuum evolution. It has been experimentally observed  in heavy ion collisions for light hadrons by the ATLAS and CMS  collaborations at the LHC~\cite{Aaboud:2018hpb,Sirunyan:2018qec} and evaluated using medium-induced corrections to the semi-inclusive fragmenting jet functions~\cite{Kang:2016ehg}.  The observables discussed in this paper,  however,  are not sensitive to the fragmentation functions in the $z \rightarrow 0$ region.  This is  because with the designed CM energies of the EIC hadrons with large transverse momentum relative to the collision axis cannot be produced with very small fragmentation fractions.

\section{Comparison with HERMES data}

In order to provide theoretical predictions for heavy flavor modification at the EIC, it is useful to get some guidance from existing DIS measurements on nuclei. The HERMES collaboration at HERA has collected such data on light hadron production, albeit at much lower center-of-mass energies.  With this limitation in mind,  we use the opportunity to test the validity of our theoretical framework of cold nuclear effects on hadronization. 
Let us define the modification of semi-inclusive pion production as follows:
\begin{equation}
R_{eA}^{\pi}(\nu,Q^2,z)=\frac{\frac{N^{\pi}(\nu,Q^2,z)}{N^e(\nu,Q^2)}\Big|_A}{\frac{N^{\pi}(\nu,Q^2,z)}{N^e(\nu,Q^2)}\Big|_D} \, , 
\label{Rhermes}
\end{equation}
where $N^{\pi}(\nu,Q^2,z)$ and $N^e(\nu,Q^2,z)$ are the event number for  hadron production  ($\pi^+$) and the total number of inelastic events determined by measuring the scattered  lepton, respectively. The kinematic variables are defined as $\nu=E-E'$, $Q^2=-(k-k')^2$, $z=E_h/\nu$, where $E(k)$ and $E'(k')$ are the energies (momenta) of the incoming and outgoing electron in the target rest frame, respectively. Subscripts $A$=Kr, Xe, ... and $D$=deuteron denote the target nuclei. The energy of incoming electrons is 27.6 GeV. Here, we employ the same kinematic cuts as in the HERMES measurements: $Q^2>1\,{\rm GeV}^2$, $W=\sqrt{2M\nu+M^2-Q^2}>2\,{\rm GeV}$ and $y=\nu/E<0.85$~\cite{Airapetian:2007vu}. The idea behind normalizing by the number of DIS events is to both account for the large number of nucleons in the nucleus and  to minimize effects strictly due to nuclear PDFs. 

Theoretically, due to the  Landau-Pomeranchuk-Migdal effect in QCD, the contribution of the in-medium shower depends on the energy  of the hard parton in the rest frame of nuclear matter, which in DIS is the scale $\nu$. For the perturbative part and PDFs, both $Q$ and $\nu$ are hard scales, especially at large Bjorken-$x$. The observables involved in this work are normalized by the inclusive cross section, and small differences due to scale choice cancel in the ratio. For the evolution, which is in the branching  momentum $k_\perp$, this scale enters just in the boundary of the allowed phase space. We have found that most of the medium shower contribution comes from $k_\perp^2 \sim$ ~1~GeV$^2$  and putting a different limit on the evolution variable, unless very small, will not affect the description of branching in matter and the extraction of its transport properties.

Figure~\ref{fig:hermes} presents comparisons between the theoretical predictions and the HERMES measurements of pion production in DIS on Kr and Xe targets. 
The bands correspond to the nuclear matter transport parameter and its variation described in the previous section,  but the splitting kernels and evolution  are for the Kr and Xe nuclei.  The theoretical predictions and HERMES data are in good agreement in a range of energy values  $\nu$, and also as a function of $z$. Pion production is more suppressed at lower $\nu$  and already hints that it will  be more beneficial to study cold matter effect at lower energy e+A collision. In addition, we can also see that there is a stronger suppression on heavier nuclear targets. As a function of $z$,  the largest attenuation is at the highest fragmentation fractions and everywhere in the studied region  $R_{eA}^{\pi}(z) < 1$.

\section{Hadron Production at the EIC}

In this section,  we  move to the main result of this work  - hadron and, especially,  heavy  meson cross section modification  at the EIC.
Here,  we consider three benchmarks energy combinations  for electron-proton collisions (for electron-nucleus collisions, the beam energy is per nucleon): 5 GeV (e) $\times$ 40 GeV (A),  10 GeV  (e) $\times$ 100 GeV (A) and 18 GeV (e) $\times$ 275 GeV (A). To investigate the nuclear medium effects, we study the ratio of the cross sections in electron-gold (e+Au) collision to the one in e+p collision.  We use the cross section of inclusive jet production for normalization that minimizes  the effect of nuclear PDFs.
\begin{equation}\label{eq:defRAatEIC}
R_{eA}^{h}(p_T,\eta,z)=\frac{\frac{N^{h}(p_T,\eta,z)}{N^{\rm inc}(p_T,\eta)}\Big|_{\rm e+ Au}}{\frac{N^{h}(p_T,\eta,z)}{N^{\rm inc}(p_T,\eta)}\Big|_{\rm e+p}} \, .
\end{equation}
Note that the kinematic variables are different than in Eq.~(\ref{Rhermes}).
Here, $N^{\rm inc}(p_T,\eta)$ denotes the cross section of large radius jet production~\cite{Li:2020rqj}  with transverse momentum  $p_T$ and rapidity $\eta$ \footnote{Here, $p_T$ is the transverse momentum relative to the electron/nulcear beam direction in the laboratory frame, which is different from the Breit frame in SIDIS. For a relativistic  particle $\eta \approx y = \ln\sqrt{(E^h+p_z^h)/(E^h-p_z^h)}$, where $E^h$ and $p_z^h$ are the energy and momentum along the beam direction, respectively, of the hadron in the laboratory frame.}.
As we only aim  to eliminate the differences  between  proton and  nuclear PDFs,  results for the inclusive jet production to lowest order are enough for this purpose. In fact, we can reasonably estimate those numbers from the number of scattered electrons in calculable $p_T$ and backward rapidity bins.

\begin{figure*}[!t]
    \centering
	\includegraphics[width=0.42\textwidth]{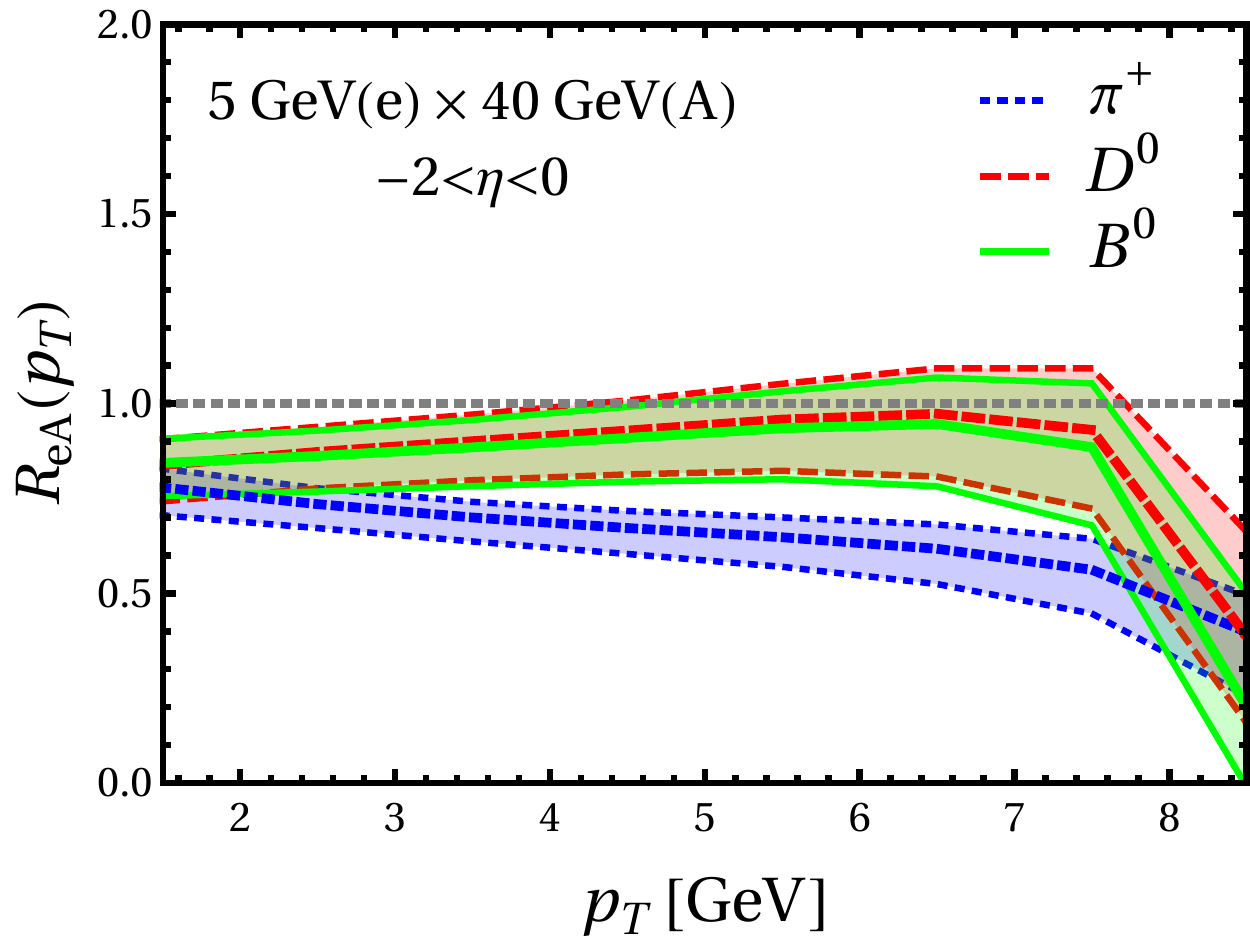} \,\,\,
	\includegraphics[width=0.42\textwidth]{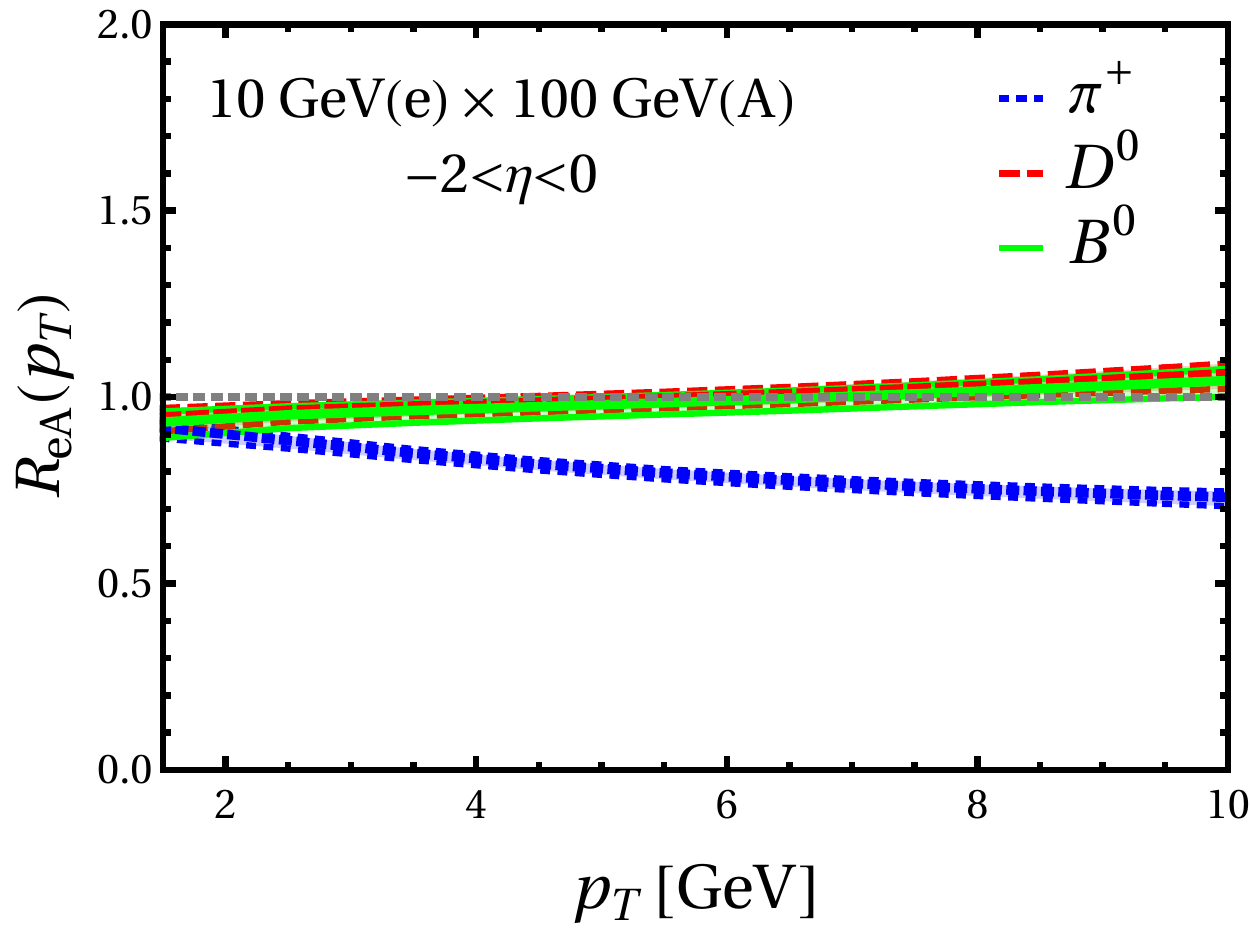}
	\includegraphics[width=0.42\textwidth]{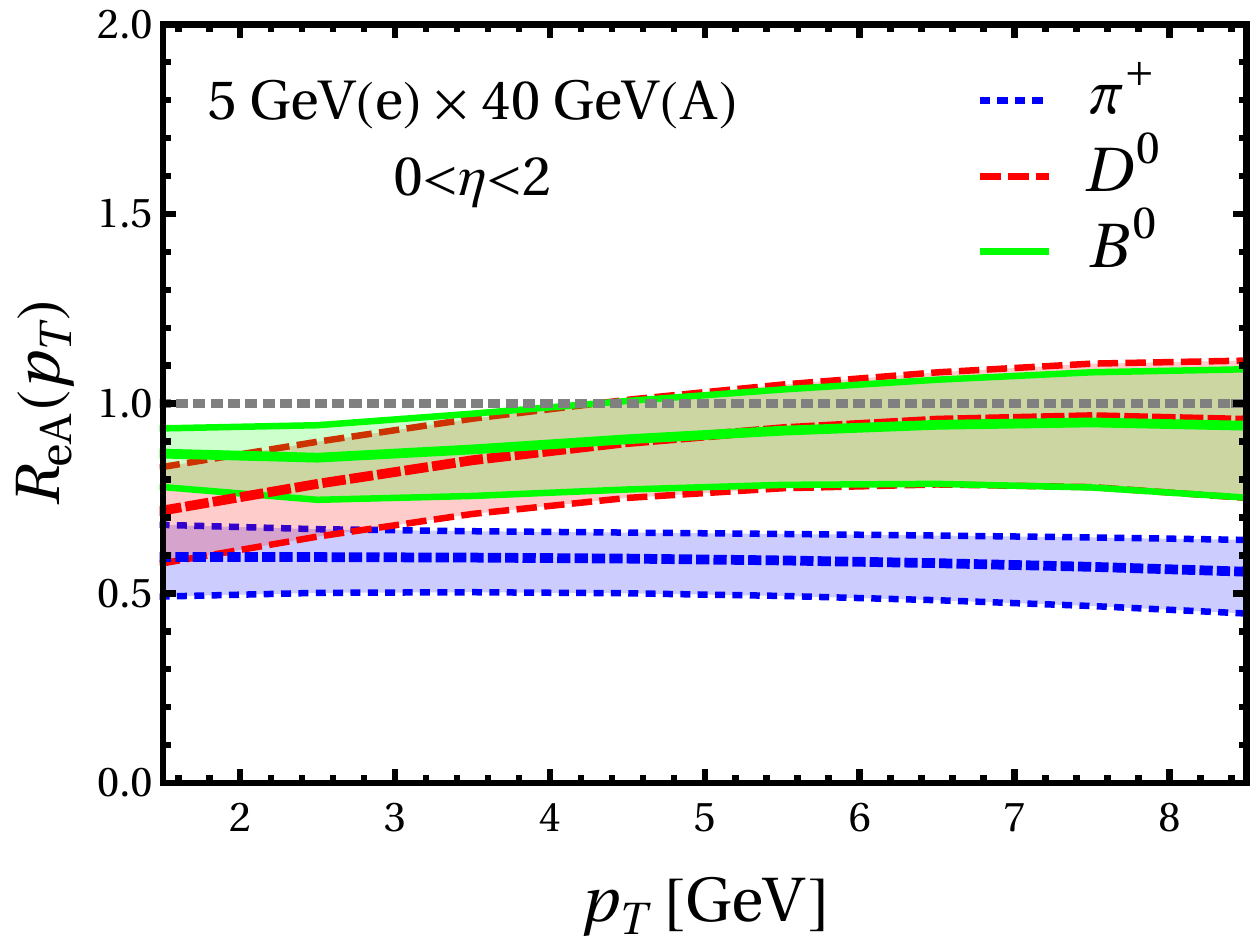}	\,\,\,
	\includegraphics[width=0.42\textwidth]{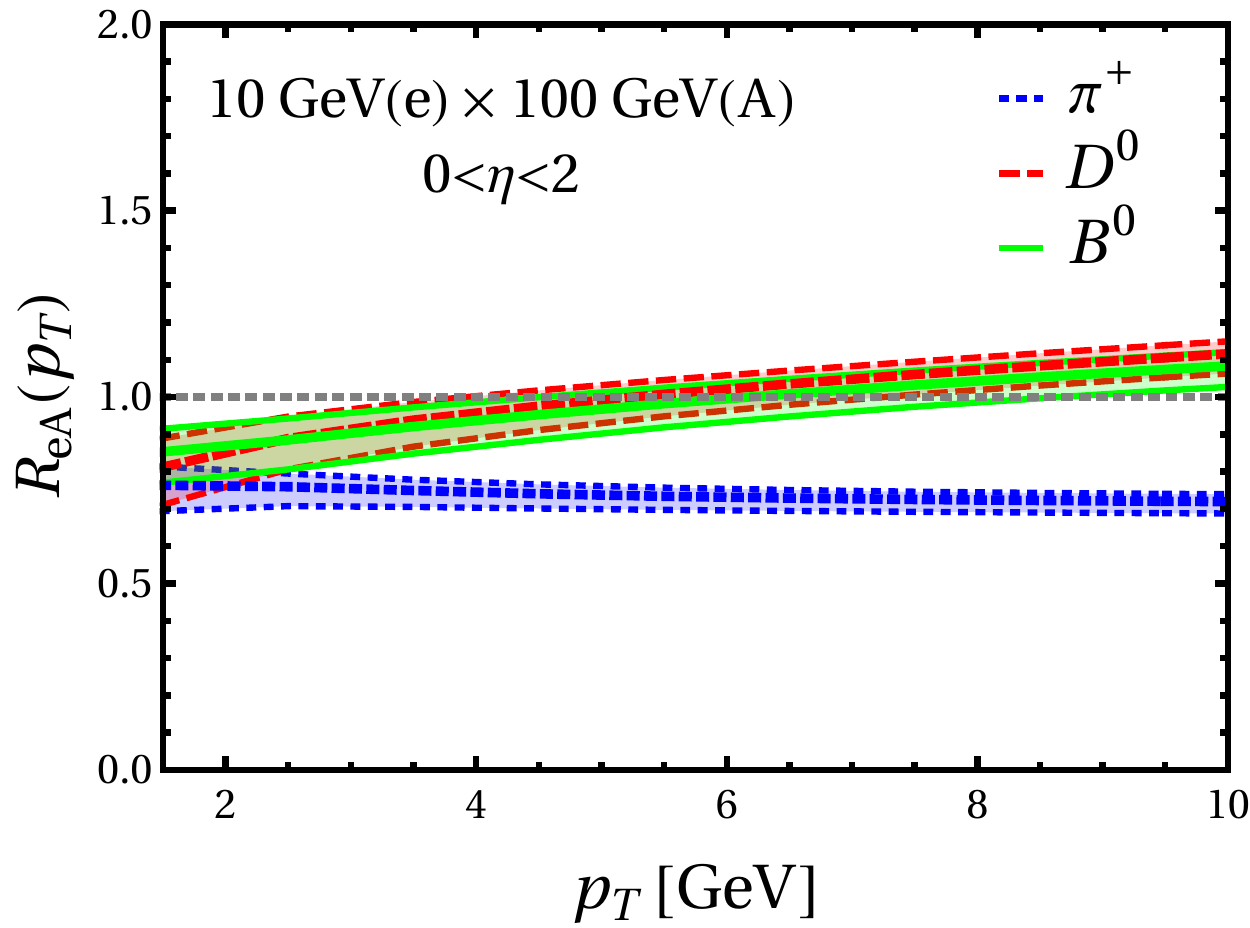}
	\includegraphics[width=0.42\textwidth]{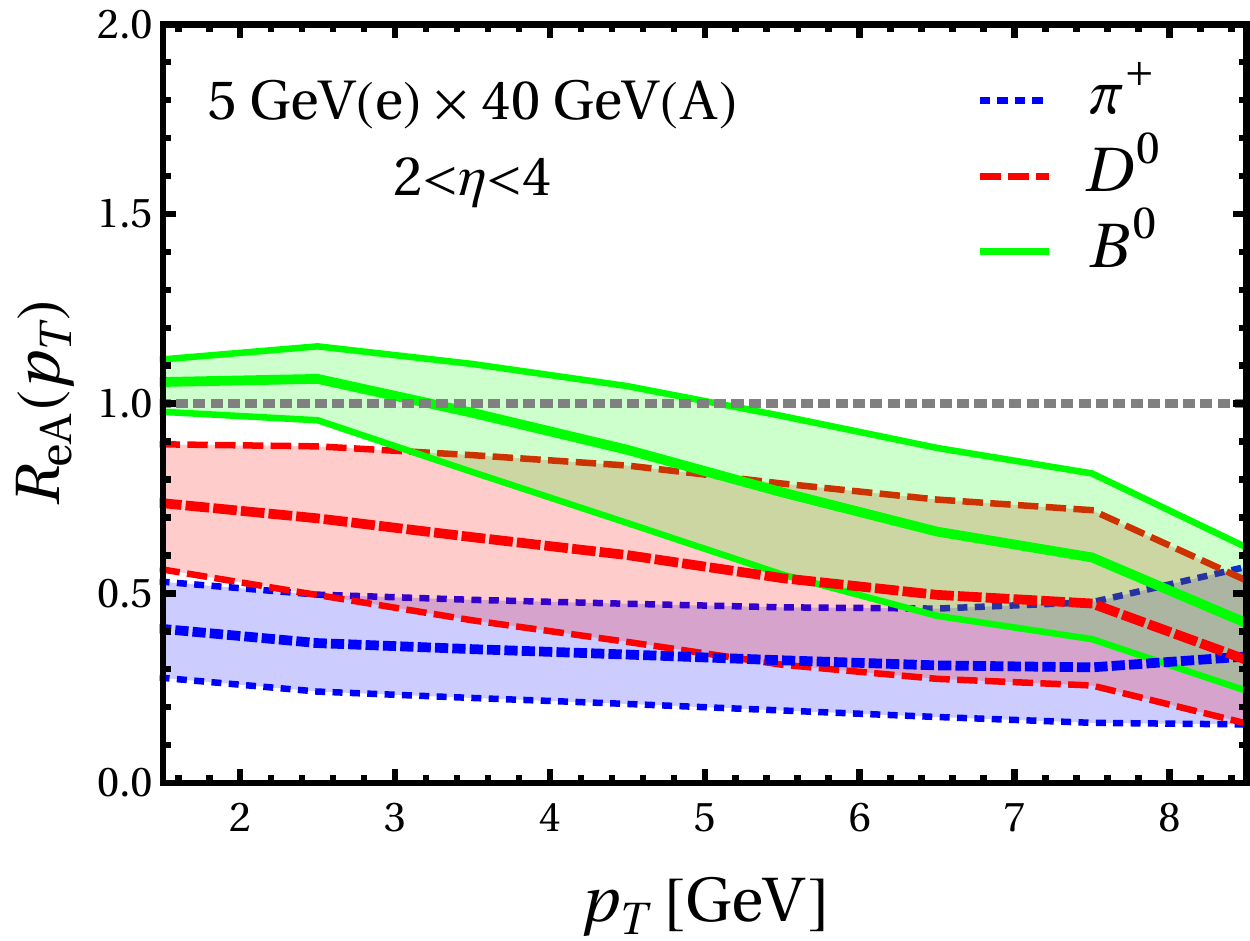}	\,\,\,
	\includegraphics[width=0.42\textwidth]{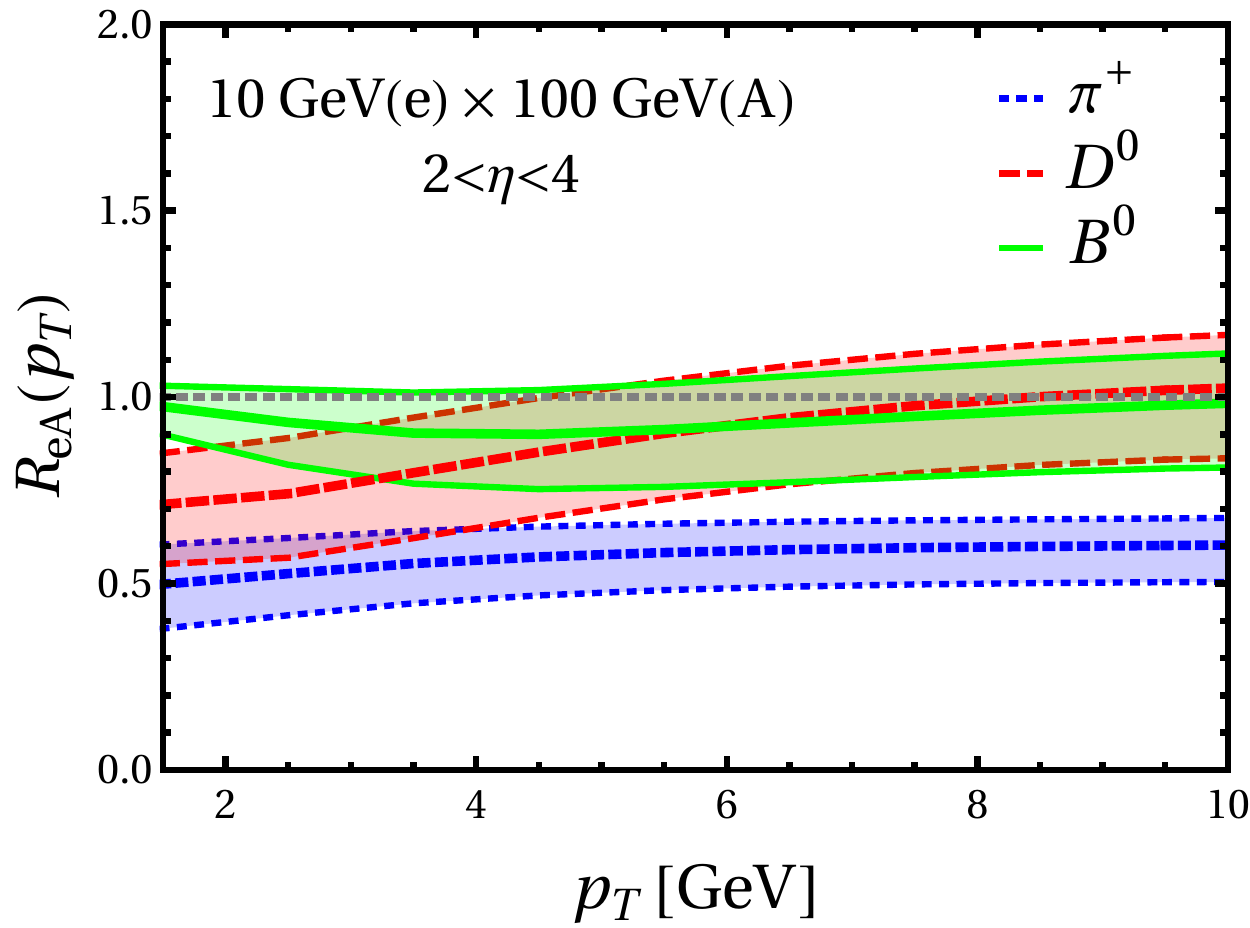}
    \caption{ Medium modification of  $\pi^+$, $D^0$ and $B^0$ production on a gold (Au) nucleus at the EIC as a function of transverse momentum in three rapidity regions for the hadrons. The left column of figures is for 5 GeV (e)  $\times$ 40 GeV (A) collisions and the right column of figures is for 10 GeV (e)  $\times$ 100 GeV (A) collisions, respectively.  The rapidity regions from top to bottom are  -2$<\eta<$0, 0$<\eta<$2 and 2$<\eta<$4.} 
 \label{fig:ptdisEIC}
\end{figure*}

We first turn  to the production of  hadrons as a function of the transverse momentum $p_T$ in the laboratory frame.
The in-medium shower corrections induced by the  interactions between the final-state parton and the nucleus vary with the parton energy in the nuclear rest frame, where the lower energy parton receives the larger medium corrections.  One way to study this effect is to vary the CM energy as shown in Fig.~\ref{fig:ptdisEIC}. The left column of panels  is for 5 GeV $\times$ 40~GeV e+Au collision  and the right column of panels  is for 10 GeV $\times$ 100 GeV ones.    The dotted blue line, dashed red line and solid green lines denote the result for $\pi^+$, $D^0$ and $B^0$, respectively.  We find that not only is the magnitude of the nuclear modification 
$R_{eA}(p_T)$ larger at the lower CM energy, but the sensitivity to the transport properties of nuclei, illustrated by the width of the theory bands, is also enhanced.  We also performed  calculations for 18 GeV $\times$ 275~GeV  e+Au collision and found that the medium effects at those energies are  smaller than the ones at 10~GeV $\times$ 100~GeV. Hence, we don't show them here.

Another way to vary the parent parton energy $\nu$ in the rest frame of the nucleus is to use different rapidity ranges. 
For given hadron $p_T$, the medium corrections will be  larger for smaller relative rapidity $|\eta - \eta_A|$, with hadron rapidity $\eta$ and nuclear rapidity $\eta_A$ in the lab frame
\footnote{In the lab frame $\eta_A\approx 4.4$ and 5.4 in 5 GeV $\times$ 40 GeV and 10 GeV $\times$ 100 GeV e+A collisions, respectively. The parent parton energy in the rest frame of the nucleus can be obtained by $\nu=p_T\cosh |\eta - \eta_A|.$ For further discussion of the differences between fixed target and collider kinematics see~\cite{Accardi:2009qv}.}. 
The horizontal sets of panels in Fig.~\ref{fig:ptdisEIC} presents $R_{eA}^{h}$ values  in three rapidity bins $-2$$<\eta<$0, 0$<\eta<$2 and 2$<\eta<$4.   The in-medium corrections are the largest in the  forward hadron rapidity region  $2<\eta<4$  as expected.  
The study of the transverse momentum distribution of hadrons  can provide a first glimpse of jet quenching effects in reactions with nuclei at the EIC.  This is especially clear for the suppression of pions. Even for heavy flavor,  at low CM energies and forward rapidities  we are beginning to see a hierarchy of suppression patterns and 
sizable suppression at large $p_T$.  At the same time,  to investigate the nature of hadronization, more differential observables are needed. This is especially true for heavy flavor which in many cases shows little or no nuclear modification. The physics reason is that away from the  edges of kinematic acceptance the range of  fragmentation fractions $z$ that give sizable contributions to hadron production  is  limited. For heavy quarks fragmenting into heavy mesons it is in the range $z = 0.55 - 0.90$ and is harder for $b$ quarks in comparison to $c$ quarks. This is precisely the range of momentum fractions where the medium-induced modification to $D$-meson and $B$-meson FFs transitions from suppression at large $z$ to enhancement at small $z$, see   Fig.~\ref{fig:FFsInMedium}. Consequently, for most energy and rapidity combinations we find 
$R_{eA}^{D} \approx R_{eA}^{B} \approx 1$.  We finally remark that since $R_{eA}$ is a ratio of cross sections,  there is practically no difference in the calculated nuclear modification with LO and NLO hard parts.

\begin{figure*}[t!]
	\centering
	\includegraphics[width=0.42\textwidth]{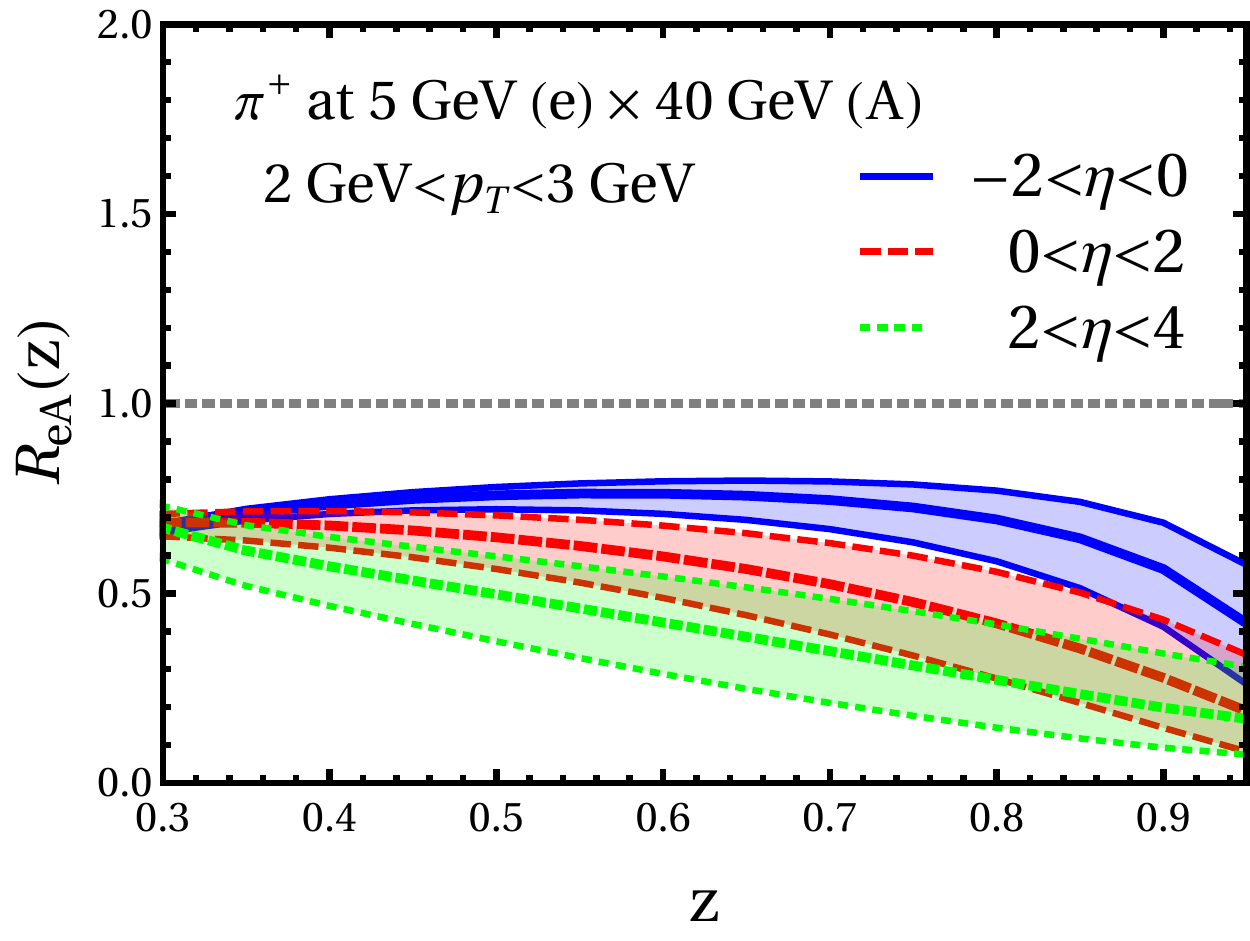}\,\,\,
	\includegraphics[width=0.42\textwidth]{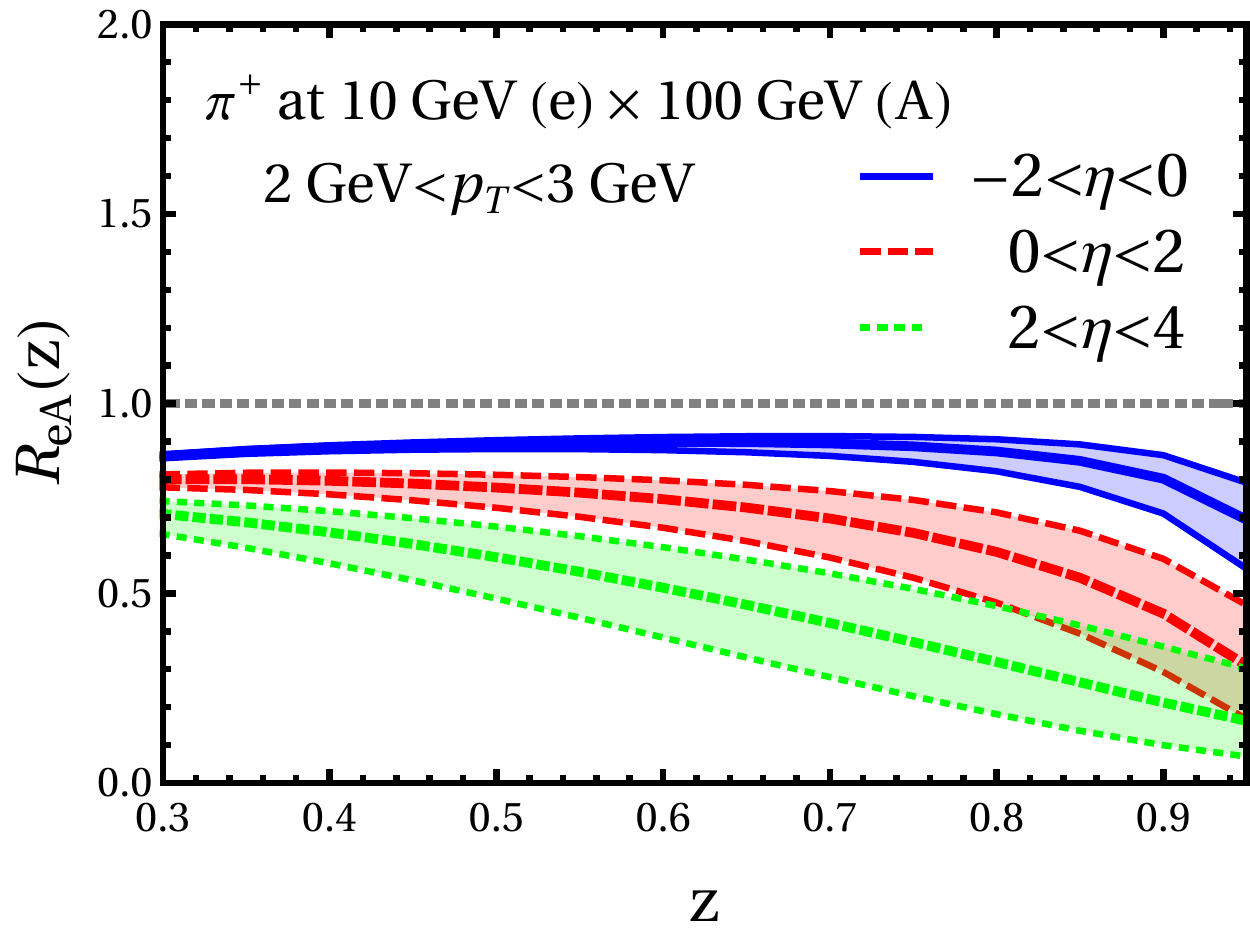}	
	\caption{ In-medium corrections for $\pi^+$ production as a function of $z$ at the EIC  in three rapidity regions. 
		Blue bands (solid lines), red bands (dashed lines), and  green bands (dotted lines) correspond to  -2$<\eta<$0, 0$<\eta<$2 and 2$<\eta<$4, respectively.
		Results for 5~GeV(e) $\times$ 40 GeV(A)  collisions are shown on the left and results for 10 GeV(e) $\times$ 100 GeV(A) collisions  are shown on the right. }
	\label{fig:zdisEIC_pi}
\end{figure*}

 \begin{figure*}[t!]
 	\centering
 	\includegraphics[width=0.42\textwidth]{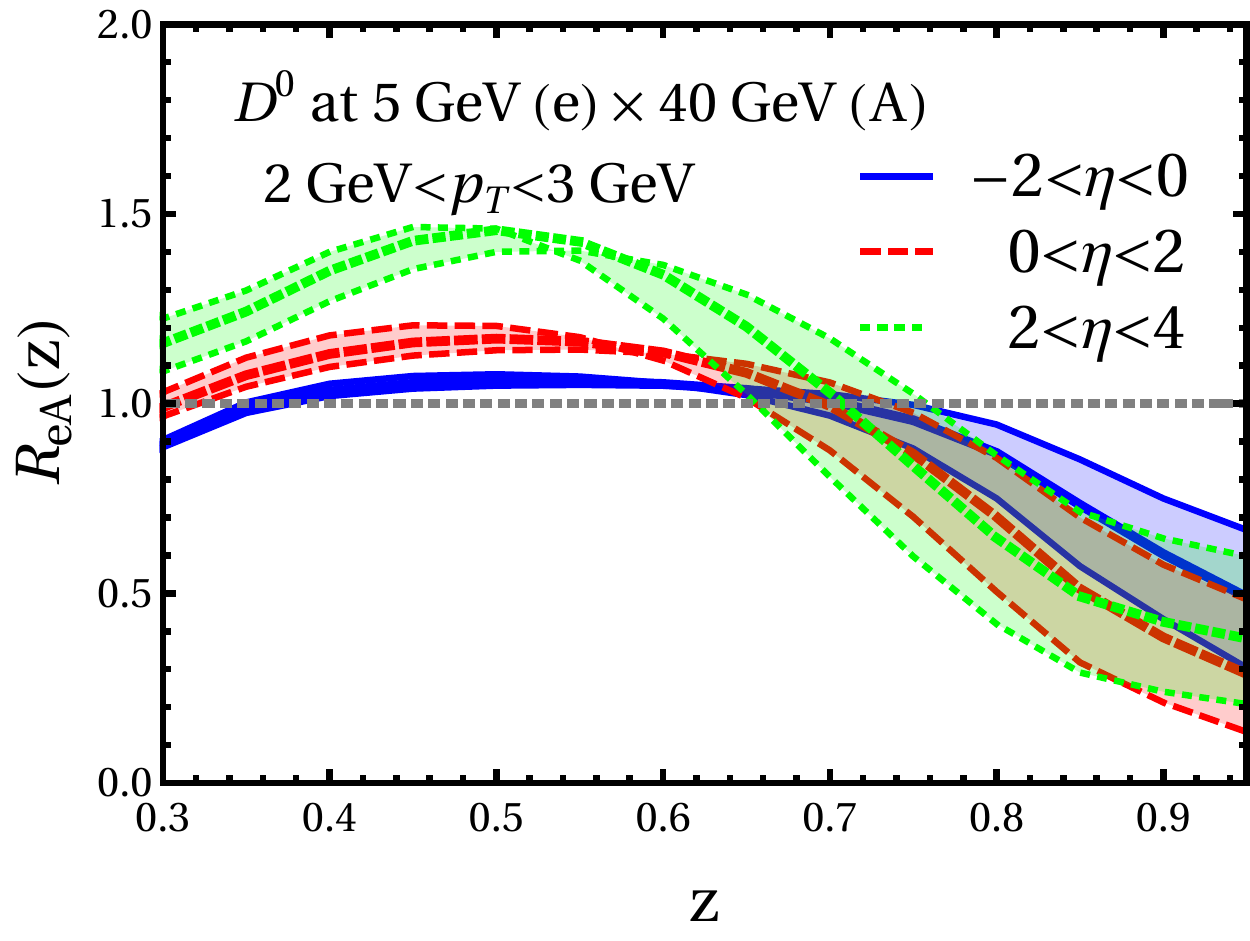}\,\,\,
 	\includegraphics[width=0.42\textwidth]{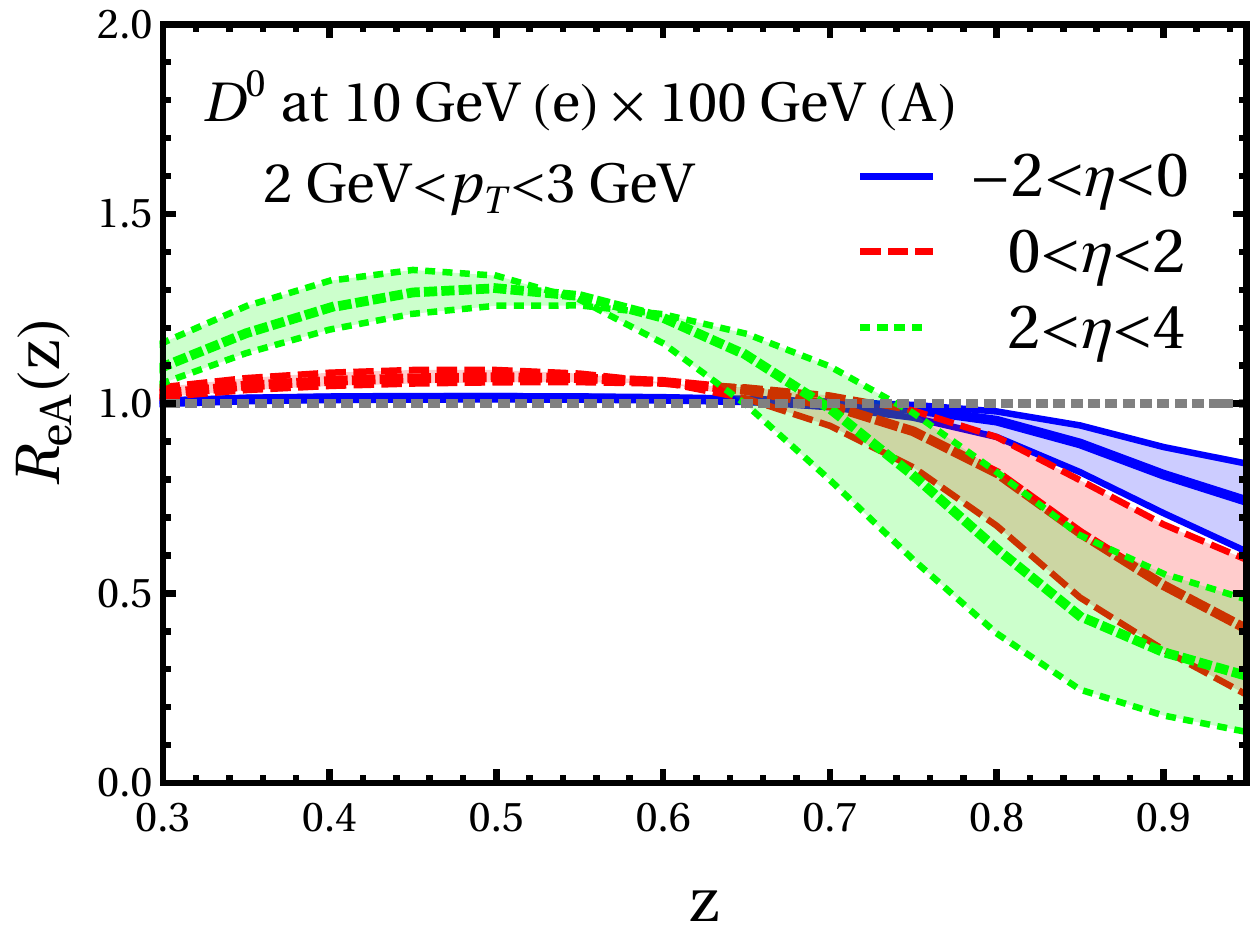}
 	\includegraphics[width=0.42\textwidth]{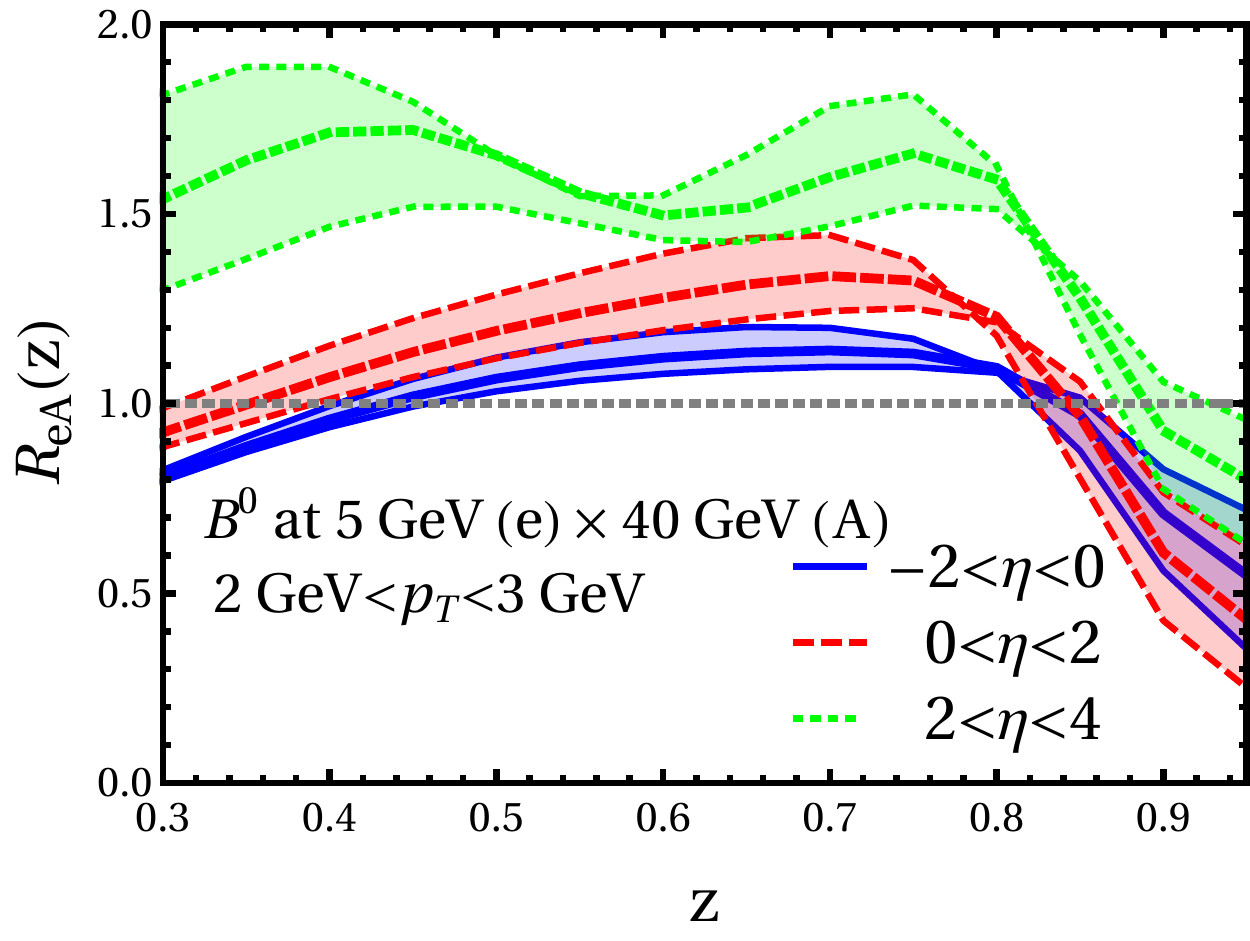}	\,\,\,
 	\includegraphics[width=0.42\textwidth]{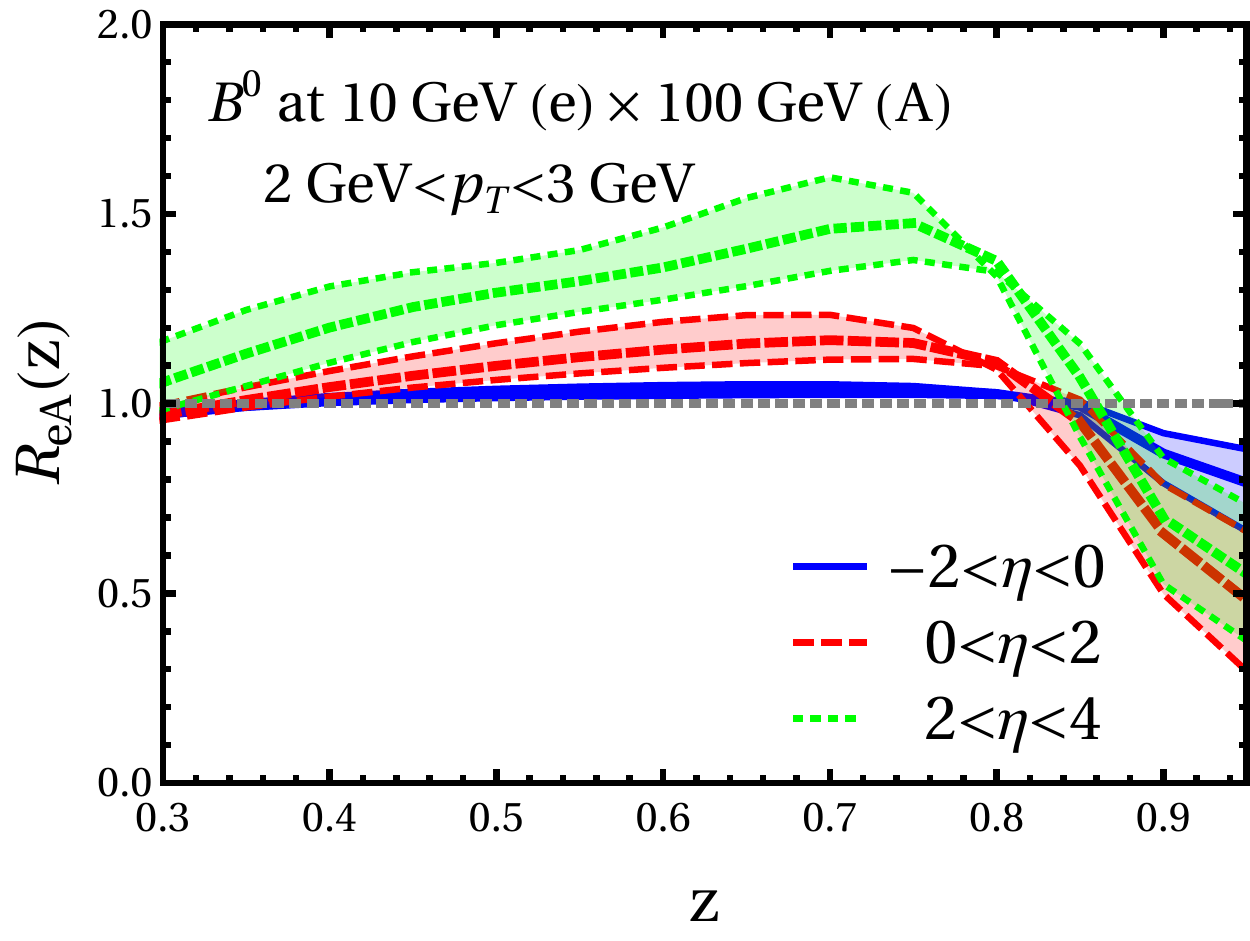}
 	\caption{ In-medium corrections for $D^0$ and $B^0$ as a function of the momentum fraction $z$ at the EIC in three rapidity regions. Top panels are for $D$-mesons and bottom panels are for $B$-mesons.  The electron and proton/nucleus beam energies,  color and line coding are the same as in  Fig.~\ref{fig:zdisEIC_pi}.}
 	\label{fig:zdisEIC_D0B0}
 \end{figure*}

To exploit the  differences in the hadronization patterns between light hadrons and heavy mesons and use them to discriminate between theoretical models of nuclear modification~\cite{Li:2020sru} we turn to more differential observables. Specifically, we fix the $p_T$ bin and 
look at the momentum fraction distribution $z$, which we extract from our calculation.  This corresponds to the variation of $\nu$ which in experiment can be constrained by the kinematics of the scattered electron.      
Figures~\ref{fig:zdisEIC_pi} and \ref{fig:zdisEIC_D0B0}  present $R_{eA}^{h}$ result as a function of $z$. 
Our  predictions in rapidity regions $-2$$<\eta<$0, 0$<\eta<$2 and 2$<\eta<$4 are represented by the blue solid lines, red dashed lines, and green dotted lines, respectively.  We choose  the $p_T$ range  2 GeV$<p_T<$3 GeV  where the cross sections at lower $z$ values (e.g. $z\sim 0.3$) are sizable.  With the same $p_T$ range fixed,  we can  identify  larger in-medium effects at 5 GeV $\times$ 40 GeV e+Au collision than at 10 GeV$\times$ 100 GeV collisions. Additionally, hadron production in the forward rapidity region $2<\eta<4$ receives the largest in-medium corrections.  For $\pi^+$ production, $R_{eA}$ is  always smaller than one in the region of momentum fractions that is accessible with the largest quenching seen at large $z$, see  Fig.~\ref{fig:zdisEIC_pi}.

In contrast to light flavor, the modification  of open heavy flavor in DIS reactions with nuclei,  such as the one for  $D^0$s  and  $B^0$s  shown in  Fig.~\ref{fig:zdisEIC_D0B0},  is much more closely related to the  details of hadronization.  The observed  $R_{eA}(z)$ is qualitatively consistent with the  effective modification of fragmentation functions seen in
Fig.~\ref{fig:FFsInMedium} even after their convolution with the PDFs and the hard part.  There is a significant suppression for large values of $z$, but it quickly evolves to      
 enhancement for $z<0.65$ and $z<0.8$ for $D$-mesons and $B$-mesons, respectively.  The effect is most pronounced at forward rapidities
 and we  find that $R_{eA}^{h}$  as a function of $z$ is a more suitable observable  for  cold nuclear  matter tomography at the EIC   than the  transverse momentum distributions' modification for hadrons in the laboratory frame alone.  We note that as we go toward small values of $z$ (e.g.  $z=0.3$)  the enhancement can be compensated by suppression
that  arises from the normalization factor $N_p^{\rm inc}(p_T,\eta)/N_A^{\rm inc}(p_T,\eta)$.  The exact interplay of these effects depends on the rapidity region of interest.

\section{Conclusions}

In summary, we presented  first predictions for heavy $D$-mesons and $B$-meson  production in e+A  collisions at the EIC including  NLO corrections and  cold nuclear matter effects.  The much higher CM energies  relative to HERMES and, correspondingly,  larger parent parton energies in the rest frame of the nucleus  boost  hadron formation times.  This motivates   a detailed theoretical study  of in-medium effects arising from final-state parton-level interactions inside large nuclei. The effective  modification of open heavy flavor fragmentation functions was obtained by solving  
 the generalized DGLAP evolution equations  with  in-medium spitting kernels derived in the framework  of SCET$_{\rm G}$.  
This theoretical approach, when applied to light hadron production,  shows good agreement with HERMES measurements and 
allows us to set a range of nuclear transport properties and make projections for the future EIC.  One should keep in mind, however,  that at HERMES energies other effects, such as hadronic rescattering, can contribute to the observed suppression of particle multiplicities.  

To demonstrate the utility of heavy flavor for cold nuclear matter tomography  we carried out a comprehensive study of the production of various  $D$-mesons and $B$-meson states at different center-of-mass energies and different rapidity ranges at the EIC.  We found that the modification of light and heavy flavor hadron cross sections in reactions with nuclei is sizable and  depends on the electron and proton/nucleus beam energy combinations and the rapidity gap between the produced hadron and the target nucleus. Our numerical results show that  the 5 GeV $\times$ 40 GeV  scenario  followed  by  the 10 GeV $\times$ 100 GeV  case and the forward  proton/nucleus going rapidity region   2$<\eta<$4  produce the largest nuclear effects. Conversely, semi-inclusive hadron production at large center-of-mass energies, e.g. 18 GeV $\times$ 275 GeV,  and  backward  rapidities, e.g. -2$<\eta<$0, exhibits only small modification in e+A reactions relative to e+p ones. Such kinematics is better suited to explore shadowing and the phenomenon of gluon saturation.    
 
Last but not least, we looked for experimental observables  that are most sensitive to the details of hadronization. While $p_T$ distributions in the laboratory frame can provide initial information on the quenching of hadrons in cold nuclear matter,  a more differential observable such as the fragmentation fraction $z$ distribution measured by HERMES is a much better choice, especially for open heavy flavor.  The clear transition from enhancement to suppression at moderate to large values of $z$ will be an unambiguous and quantitative measure of parton shower formation in large nuclei.  In conclusion, we expect that this work will be useful in  guiding the future light and  heavy flavor tomography program at the EIC.

\section*{Acknowledgments}
This work was supported by the U.S. Department of Energy under Contract No. DE-AC52-06NA25396, the Los Alamos National Laboratory LDRD program, and the TMD topical collaboration for nuclear theory.

\bibliographystyle{JHEP}
\bibliography{mybib}

\end{document}